\documentclass [winedt,yap]{iitparc}
\usepackage{cite}
\usepackage{amsmath,amssymb,amsfonts,bm}
\usepackage{eufrak}
\usepackage{lscape}
\usepackage{floatfig,wrapfig,epsfig}
\usepackage{subfigure}
\usepackage{color}
\usepackage{psboxit}
\usepackage{rotating}
\usepackage{curves}
\usepackage{ulem}
\usepackage[mathscr]{eucal}
\RequirePackage{psfrag}
\RequirePackage{graphicx}

\usepackage{amsxtra}

\begin{document}\normalem
\initfloatingfigs
\frontmatter          

\IssuePrice{25.00}%
\TransYearOfIssue{2012}%
\TransCopyrightYear{2012}%
\OrigYearOfIssue{2012}%
\OrigCopyrightYear{2012}%

\TransVolumeNo{73}%
\TransIssueNo{1}%
\OrigIssueNo{1}%


\mainmatter

\setcounter{page}{161}
\CRubrika{MULTI-AGENT SYSTEMS}
\Rubrika{MULTI-AGENT SYSTEMS}

%

\title{A Cyclic Representation of Discrete Coordination Procedures\thanks{This work was supported in part by the Russian
Foundation for Basic Research, project no.~09-07-00371 and the program of RAS Presidium ``Mathematical Theory of Control.''}}

\author{R. P. Agaev and P. Yu. Chebotarev}

\institute{Trapeznikov Institute of Control Sciences, Russian Academy of Sciences, Moscow, Russia}

\titlerunning{A Cyclic Representation of Discrete Coordination Procedures}

\authorrunning{Agaev, Chebotarev}

\OrigCopyrightedAuthors{R.P. Agaev and P.Yu. Chebotarev}

\received{Received May 20, 2011}

\OrigPages{pp.~178--183}

\maketitle

\def\rank{\mathop{{\rm rank}}\nolimits}          
\def\R{\mathbb{R}}                               
\def\C{\mathbb{C}}                               
\def\ind{\mathop{\rm ind}\nolimits}              
\def\NN{\mathop{\mathcal N}\nolimits}            
\def\RR{\mathop{\mathcal R}\nolimits}            
\def\sqa{\sqcap\!\!\!\!\sqcup}                   
\def\epr{\hfill$\sqa$\smallskip}                 
\def\l{\ell}                                     
\def\Pbes{P^\infty}                              
\def\aa{\alpha}                                  
\def\e{\varepsilon}                              
\def\G{\Gamma}                                   
\def\bb{\beta}                                   
\def\si{\sigma}                                  
\def\cdc{,\ldots,}                               
\def\1n{1,\ldots,n}                              
\def\J{\tilde{J}}                                
\def\ms{\mathstrut}                              
\def\xz{\hspace{-.07em}}                         
\def\xy{\hspace{.07em}}                          
\def\tr{\operatorname{tr}}                       
\accentedsymbol{\Pin}{\stackrel{\;\scriptscriptstyle\infty}{P_{\mathstrut}}} 
\def\Up#1{\vspace{-#1em}}                        
\def\beq{\begin{equation}}                       
\def\eeq{\end{equation}}                         
\def\eq#1{\begin{equation}#1\end{equation}}                     
\def\O{0}                                        
\def\B{{\protect\mathfrak B}}                    
\def\suml {\mathop{\sum}   \limits}              
\def\prodl{\mathop{\prod} \limits}               

\newlength{\widebarargwidth}
\newlength{\widebarwidth}
\newlength{\widebarargheight}
\newlength{\widebarargdepth}
\DeclareRobustCommand{\wbar}[1]{%
  \settowidth{\widebarargwidth}{\ensuremath{#1}}%
  \settoheight{\widebarargheight}{\ensuremath{#1}}%
  \settodepth{\widebarargdepth}{\ensuremath{#1}}%
  \addtolength{\widebarargwidth}{-0.7\widebarargheight}
  \addtolength{\widebarargwidth}{-3.8\widebarargdepth}
  \makebox[0pt][l]{\addtolength{\widebarargheight}{-0.2ex}
    \hspace{0.2\widebarargheight}%
    \hspace{0.2\widebarargdepth}%
    \hspace{0.5\widebarargwidth}%
    \setlength{\widebarwidth}{0.6\widebarargwidth}%
    \addtolength{\widebarwidth}{0.6ex}
    \makebox[0pt][c]{\rule[\widebarargheight]{\widebarwidth}{0.1ex}}}
  {#1}}

\begin{abstract}
We show that any discrete opinion pooling procedure with positive weights can be asymptotically approximated by DeGroot's procedure whose communication digraph is a Hamiltonian cycle with loops. In this cycle, the weight of each arc (which is not a loop) is inversely proportional to the influence of the agent the arc leads to.
\end{abstract}

\section{INTRODUCTION}

One of the first discrete models of reaching consensus (decentralized coordination) was proposed by DeGroot~\cite{DeGroot74}.
Suppose that $s(0)=(s_1^0\cdc s_n^0)^{\rm T}$ is the vector of initial opinions of the members of a group and ${s(k)=(s_1^k\cdc s_n^k)^{\rm T}}$ is the vector of their opinions after the $k$th step of coordination. In accordance with DeGroot's model, $s(k)=Ps(k-1),\;k=1,2,\ldots,$ where $P$ is a row stochastic influence matrix whose entry $p_{ij}$ specifies the degree of influence of agent $j$ on the opinion of agent~$i.$
Thereby,
\eq{
\label{281110eq1}
s(k)=P^ks(0),\quad k=1,2,\ldots.
}

\emph{A consensus is\/} [asymptotically] \emph{reached} if $\lim_{k\to\infty}s_i^k=\bar s$ for some $\bar s\in\R$ and all $i\in\{\1n\}.$ {De\-Groot} states that a~consensus is reached for any initial opinions if and only if the matrix ${\Pbes=\lim_{k\to\infty}P^k}$ exists and all rows of $\Pbes$ are identical, which is equivalent to the regularity\footnote{A stochastic matrix is said to be \emph{regular\/} if it has no eigenvalues of modulus~1 except for the simple eigenvalue~$1.$} of~$P$.

If $P$ is not regular, then the opinions do not generally tend to agreement. Yet, a consensus can be reached if the vector of initial opinions belongs to a certain subspace. In~\cite{AgaChe11}, we give a characterization of this subspace and propose the \emph{method of orthogonal projection} which generalizes DeGroot's method. It is shown that in the method of orthogonal projection, as well as in DeGroot's method, no nonbasic agent can affect the final result.

In this paper, we show that the result of any DeGroot's procedure with a strong communication digraph as well as the result of any procedure of orthogonal projection without nonbasic agents can be represented by DeGroot's procedure whose communication digraph is a Hamiltonian cycle with loops. In this cycle, the weight of each arc (which is not a loop) is inversely proportional to the influence of the agent the arc leads to.

\section{BASIC CONCEPTS AND RESULTS}
\label{s_Notat}

With a stochastic influence matrix $P$ in DeGroot's model we associate the \emph{communication digraph\/} $\G$ with vertex set $V(\G)=\{\1n\}.$ $\G$~has a $(j,i)$ arc with weight $w_{\xz ji}=p_{ij}$ whenever $p_{ij}>0$ (i.e., whenever agent $j$ influences agent~$i$). Thus, arcs in $\G$ are oriented \emph{in the direction of influence}; the weight of an arc is the power of influence.

The \emph{Kirchhoff matrix\/} (see \cite{Tutte84,CheAga02ap}) $L=L(\G)=(\l_{ij})$ of digraph $\G$ is defined as follows: if $j\ne i,\,$ then $\l_{ij}=-w_{\xz ji}$ whenever $\G$ has the $(j,i)$ arc and $\l_{ij}=0$ otherwise; $\l_{ii}=\suml_{k\ne i}w_{ki},\,$ $i={\1n}$.
$I$~will denote the identity matrix of appropriate dimension.

By virtue of the above definitions, for the digraph $\G$ associated with $P$ we have
\eq{
\label{e_L:I-P}
L(\G)=I-P.
}

Any maximal by inclusion strong (i.e., with mutually reachable vertices) subgraph of a digraph is called a {\it strong component} (or a {\it bicomponent}) of this digraph. A {\it basic bicomponent\/} is a bicomponent such that the digraph has no arcs coming into this bicomponent from outside. Vertices belonging and not belonging to basic bicomponents are called {\it basic\/} and {\it nonbasic\/}, respectively. Similarly, we call an agent {\it basic}/{\it nonbasic} whenever the vertex representing this agent is {\it basic}/{\it nonbasic}. Let $b$ and $\nu$ be the number of basic vertices and the number of basic bicomponents in $\G,$ respectively.

If a consensus in DeGroot's model is reached and vertex $j$ is nonbasic, then, as stated in \cite{DeGroot74}, column $j$ in the limiting matrix $\Pbes$ is zero and the initial opinion of agent $j$ does not affect the resulting opinion.

Now we present the results from algebraic graph theory used in this paper. A longer list of results useful for decentralized control is given in~\cite{AgaChe11}.

If the sequence of powers $P^k$ of a stochastic matrix $P$ has a limit $\Pbes,$ then
\eq{
\label{e_Pbes_L}
\Pbes=\J,
}
where $\J$ is the normalized matrix of maximum out-forests of the corresponding weighted digraph~$\G$ (a corollary of the matrix tree theorem for Markov chains~\cite{WentzellFreidlin70a}).

$\Pbes$ is the eigenprojection corresponding to $0$ (principal idempotent) of $L$ and
\eq{
\label{e_ranks}
\rank\Pbes=\nu;\quad\rank L=n-\nu,
}
where $\nu$ is the number of basic bicomponents in~$\G$ \cite[Proposition~11]{AgaChe00}.
By\:\eqref{e_ranks}
$\dim\NN(L)=\nu,$
where $\dim\NN(L)$ is the dimension of the kernel (the \emph{nullity}) of~$L.$ Finally, by \cite[Proposition~12]{CheAga02ap}, $\ind L=1,$
where $\ind L$ (the index of $L$) is the order of the largest Jordan block of $L$ corresponding to the zero eigenvalue. This implies that
\eq{
\label{e_mult0}
m_L(0)=\nu,
}
where $m_L(0)$ is the multiplicity of $0$ as an eigenvalue of~$L$.

\section{CONVERGENCE IN DEGROOT'S MODEL AND PROPERTIES OF THE COMMUNICATION DIGRAPH}
\label{s_Deso}

As noted above, DeGroot's method with matrix $P$ leads to a consensus for any initial opinions if and only if there exists a limiting matrix ${\Pbes=\lim_{k\to\infty}P^k}$ and all its rows are identical.
The equality of the rows of $\Pbes$ implies that $\Pbes=\bm1\pi^{\rm T}$ for some
probability vector (the components are non-negative and sum to~$1$)~$\pi,$ where
$\bm1=(1\cdc 1)^{\rm T}.$ In this case, the consensus $\bar s$ is expressed by the inner product of the vectors $\pi$ and~$s(0)$:
\eq{
\label{e_pis}
s(\infty)=\Pbes s(0)=\bm1\pi^{\rm T}s(0)=\bm1\bar s,
}
where $s(\infty)$ is the resulting vector of opinions, $\pi$ is the final weight distribution of the DeGroot algorithm, and ${\bar s=\pi^{\rm T}s(0)}$ is the consensus.

A probability vector $\pi$ is called a {\it stationary vector} of a stochastic matrix $P$ if it is a left eigenvector of $P$ corresponding to the eigenvalue\:$1$: $\pi^{\rm T}P=\pi^{\rm T}$.
Obviously, this condition is satisfied for the vector $\pi$ in the representation $\Pbes=\bm1\pi^{\rm T}$ of $\Pbes,$ provided that the convergence of DeGroot's method is guaranteed by the regularity of~$P.$

By Theorem~3 in \cite{DeGroot74}, if for any vector of initial opinions $s(0),$ DeGroot's method converges to the consensus $\pi^{\rm T}s(0),$ then
$\pi$ is a \emph{unique\/} stationary vector of~$P$.

Let us formulate a criterion of convergence in DeGroot's model in terms of the communication digraph~$\G.$
The equality of the rows of $\Pbes$ is equivalent to $\,\rank\Pbes=1.$ Therefore, owing to \eqref{e_ranks}, when the sequence $P^k$ converges,  consensus is reached for any initial opinions if and only if the communication digraph $\G$ corresponding to $P$ has a single basic bicomponent ($\nu=1$). Consequently, provided that the sequence $P^k$ converges, $\nu=1$ is equivalent to the regularity of~$P$. In turn, by \eqref{e_mult0}, this is the case if and only if $0$ is a simple eigenvalue of~$L.$

Finally, $\nu=1$ if and only if $\G$ has a spanning \emph{out-tree} (also called \emph{arborescence\/} and \emph{branching}) \cite[Proposition\:6]{AgaChe00}.
In this case (see\,\eqref{e_Pbes_L}), $\Pbes\!=\!(p_{\ms ij}^{\ms\scriptscriptstyle\infty})\!=\!\J\!=\!(\J_{ij})$ is the normalized matrix of spanning out-trees:
\eq{
\label{e_J1}
p_{ij}^{\ms\scriptscriptstyle\infty}=\pi_j=\J_{ij}=\frac{t_j}t,\quad i,j=\1n,
}
where $t_j$ is the total weight\footnote{The weight of an out-tree (and, more generally, of a digraph) is the product of the weights of all its arcs.} of $\G$'s spanning out-trees rooted at $j\,$ and $t$ is the total weight of all spanning out-trees of~$\G.$

A survey of some results on DeGroot's iterative pooling model and its generalizations can be found in~\cite{Jackson08,AgaChe10UBSE,Gilardoni93}. Note that one of the new applications of DeGroot's model is information control in social networks~\cite{BarabanovKorNovChk10}.

\section{Representing consensus procedures by weighted Hamiltonian cycles with loops}
\label{s_Hamil}

As shown in \cite[Section\:6]{AgaChe11}, the final result of the method of orthogonal projection can be represented by a weight vector~$\aa$: the inner product of $\aa$ and the vector of initial opinions gives the consensus: ${\bar s=\aa^{\rm T}s(0)}$. This is analogous to the representation of the result of a convergent DeGroot's method: $\Pbes=\bm1\pi^{\rm T}$ and $\,{\bar s=\pi^{\rm T}s(0)}$ (see\,\eqref{e_pis}).

On the other hand, given a probability vector $\pi,$ it is easy to construct a weighted communication digraph generating $\pi$ as the final weight distribution (the stationary vector of~$P$) of DeGroot's method.

In this section, we show that all positive weight distributions on the set of agents' opinions are generated by a rather narrow class of digraphs, namely, Hamiltonian cycles of the form $n\to (n-1)\to\cdots\to 2\to 1\to n$ with loops, where the vertices are denoted by $\1n.$

By \eqref{e_Pbes_L} $\Pbes=\J,$ where $\J$ is the normalized matrix of maximum out-forests of the weighted digraph~$\G$ corresponding to~$P.\,$ Thus, given $\pi,$ a necessary and sufficient condition of the fulfilment of $\Pbes=\bm1\pi^{\rm T}$ is
\eq{
\label{e_ha1}
\J=\bm1\pi^{\rm T}.
}

A Hamiltonian cycle is a strong digraph, so its maximum out-forests are precisely spanning out-trees (see~\eqref{e_J1}). Recall that the normalized matrix of spanning out-trees (which in this case coincides with~$\J$) is the matrix whose $(i,j)$-entry is $t_j/t,$ $i,j=\1n$. Adding loops does not change the set of out-trees.

Thus, to solve the problem, i.e., to implement a given probability vector $\pi\!>\!0$ as the final weight distribution of DeGroot's method with a communication digraph in the form if a Hamiltonian cycle with loops, it is sufficient to construct a weighted cycle whose vector $(t_1\cdc t_n)^{\rm T}$ is proportional to~$\pi.$ Indeed, in this case $(t_1/t\cdc t_n/t)^{\rm T}$ coincides with $\pi,$ which guarantees~\eqref{e_ha1}. Such a cycle can be constructed by means of the following lemma.

\begin{lemma}
\label{250810th2}
For any positive vector $q=(q_{1}\cdc q_{n})^{\rm T},$ there exists a unique weighted Hamiltonian cycle of the form $n\to(n-1)\to\cdots\to 2\to 1\to n$ whose vector $(t_{1}\cdc t_{n})^{\rm T}$ of total weights of out-trees coincides with~$q$. The weight of the arc entering vertex $k$ in this cycle is $\,q_k^{-1}\!\cdot\!\!\sqrt[n-1]{\prod_{i=1}^{n\phantom{i}}q_i},\,$ $k=\1n.$
\end{lemma}

The proof of Lemma\:\ref{250810th2} is given in the Appendix.
Now we apply Lemma\:\ref{250810th2} to solve our problem.

\begin{proposition}
\label{p_picy}
For any positive probability vector $\pi=(\pi_1\cdc\pi_n)^{\rm T},$ there exists a family of weighted Hamiltonian cycles of the form
$n\to(n-1)\to\cdots\to1\to n$ with loops such that using each of them as the communication digraph in DeGroot's method implements the consensus ${\bar s=\pi^{\rm T}s(0)}$ for any vector of initial opinions~$s(0)$. The weight of the arc entering vertex $k$ in such a cycle is proportional to~$\pi_k^{-1}.$
\end{proposition}

As stated above, any communication digraph whose vector $(t_1\cdc t_n)$ of the weights of out-trees is proportional to $\pi$ implements $\pi$ as the final weight distribution of DeGroot's method. Therefore, to prove Proposition\:\ref{p_picy}, it is sufficient to observe that by Lemma\:\ref{250810th2}, a Hamiltonian cycle with loops and any given order of visiting vertices can be taken as such a digraph. The only thing that should be taken care of is that this cycle must be a \emph{communication\/} digraph, i.e., the matrix $I-L$ (see\:\eqref{e_L:I-P}) corresponding to it must be stochastic. Since $L$ has zero row sums and nonpositive off-diagonal entries, $I-L$ is stochastic if and only if the diagonal entries of $L$ are less than or equal to~$1.$ For a Hamiltonian cycle with loops, this condition is satisfied if and only if all its arc weights do not exceed~$1.$

\begin{example}\label{ex_Ham2}
{\rm Consider the influence matrix
\[
P=\left(
  \begin{array}{rrrr}
0.9  &0.1  &0    & 0   \\
0    &0.75 &0.25 & 0   \\
0.25 &0.3  &0.1  & 0.35\\
0.2  &0.15 &0    & 0.65\\
 \end{array}
\right)
\]
and the corresponding communication digraph (Fig.\:\ref{f_011211}a, which does not show loops for simplicity).

\begin{figure}[t]
\centering{\includegraphics[height=1.7in]{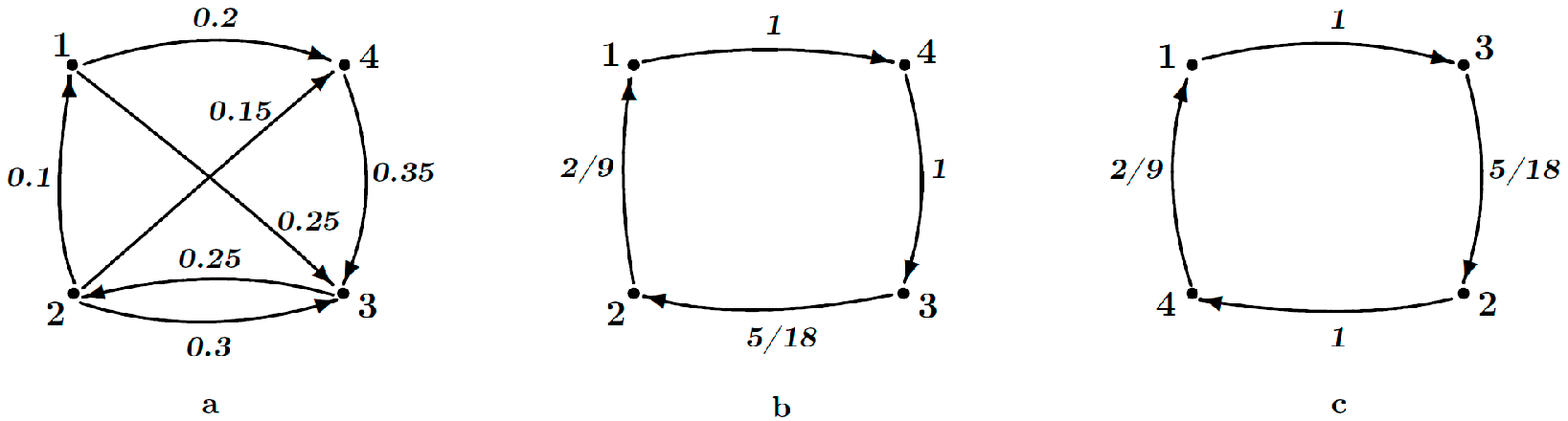}}
\caption{\label{f_011211}}
\end{figure}

Proposition\:\ref{p_picy} enables one to construct another communication digraph, in the form of a Hamiltonian cycle with loops,
that shares the final consensus obtained through DeGroot's method with the given digraph for any initial opinions.

Observe that $P$ is regular and its stationary vector is: $\pi\!=\!\frac{1}{101}(45,36,10,10)^{\rm T}.$
As stated in Section\:\ref{s_Deso}, $\Pbes\!=\!\bm1\pi^{\rm T}.$
Now we construct a Hamiltonian cycle $H$ with loops such that $P_H^{\infty}=P^{\infty}=\J$.
Proposition\:\ref{p_picy} implies that the vector of arc weights indexed by the vertices of $H$ can be $x=101\bb(\frac1{45},\,\frac1{36},\,\frac1{10},\,\frac1{10})^{\rm T},$ where $0<\bb\le\frac{10}{101}.$ Taking $\bb=\frac{10}{101}$ we obtain $x=(\frac29,\,\frac5{18},\,1,\,1)^{\rm T}.$
In this communication digraph (which is shown in Fig.\:\ref{f_011211}b), loops with weights $\frac79$ and $\frac{13}{18}$ must be attached to vertices $1$ and $2,$ respectively.
It follows from Proposition\:\ref{250810th2} that if the weights of arcs entering each vertex are preserved, then the order of vertices in the cycle can be arbitrary. In particular, the cycle in Fig.\:\ref{f_011211}c is also suitable.

Thus, for each cycle with loops of this type, DeGroot's method leads to the same final consensus procedure (determined by the weight distribution) as for the original communication digraph.
}
\end{example}

Since in Proposition\:\ref{p_picy}, $\pi_i$ is the weight of the original opinion of $i$th agent in the final consensus (the ``influence'' of the $i$th agent), it is worth noting that for communication digraphs in the form of Hamiltonian cycles with loops, this weight is inversely proportional to the weight of the arc entering vertex~$i.$ Thus, in the upshot, the most powerful agent is the one least subject to the influence of the previous agent in the cycle rather than the one maximally affecting the next agent.

Finally, Proposition\:\ref{p_picy} enables one, for any given coordination procedure based on the method of orthogonal projection~\cite{AgaChe11}, to construct a DeGroot algorithm (with the communication digraph in the form of a Hamiltonian cycle) that leads to the same consensus for any initial opinions.

\section{CONCLUSION}

It is shown that any convergent discrete iterative pooling procedure taking into account the opinions of all agents with positive weights can be approximated (in terms of achieving the same end result) by DeGroot's procedure whose communication digraph is a Hamiltonian cycle with loops. The weight of the arc entering vertex $i$ in this cycle is inversely proportional to the influence of agent~$i,$ while the order of visiting vertices can be arbitrary.

\appendix{}

\PLE{\ref{250810th2}}
For any vertex in a Hamiltonian cycle, there is exactly one tree outgoing from this vertex. That tree includes all arcs of the cycle except for the arc entering this vertex. Hence the elements of the vector $(t_1\cdc t_n)^{\rm T}$ of total weights of out-trees are given by the equations
\eq{
\label{e_tthx}
t_k
=\prod_{j\ne k}x_j
=x_k^{-1}\prodl_{j=1}^nx_j,\quad k=\1n,
}
where $x_j$ is the weight of the arc entering vertex~$j.$ Consequently,
\begin{gather*}
x_k
=t_k^{-1}\prod_{j=1}^nx_j
=t_k^{-1}\sqrt[n-1]{\prod_{i=1}^n\prod_{j\ne i}x_j}
=\frac{\sqrt[n-1]{\prod_{i=1}^{n\phantom{i}}t_i}}{t_k},\quad k=\1n.
\end{gather*}

Thus, for the fulfillment of $t_k=q_k\;(k=\1n)$ it is necessary that
\eq{
\label{e_xthq}
x_k
=\frac{\sqrt[n-1]{\prod_{i=1}^{n\phantom{i}}q_i}}{q_k},\quad k=\1n.
}

The sufficiency of \eqref{e_xthq} is verified by substituting \eqref{e_xthq} into~\eqref{e_tthx}.
The lemma is proved.
\epr


\end{document}